\documentclass{PoS}

\usepackage{epsfig}
\usepackage{axodraw}

\catcode`\@=11
\def\@citex[#1]#2{\if@filesw\immediate\write\@auxout{\string\citation{#2}}\fi
  \def\@citea{}\@cite{\@for\@citeb:=#2\do
    {\@citea\def\@citea{,\penalty\@m}\@ifundefined
       {b@\@citeb}{{\bf ?}\@warning
       {Citation `\@citeb' on page \thepage \space undefined}}%
\hbox{\csname b@\@citeb\endcsname}}}{#1}}

\def\citer{\@ifnextchar [{\@tempswatrue\@citexr}{\@tempswafalse\@citexr[]}}
\def\@citexr[#1]#2{\if@filesw\immediate\write\@auxout{\string\citation{#2}}\fi
  \def\@citea{}\@cite{\@for\@citeb:=#2\do
    {\@citea\def\@citea{--\penalty\@m}\@ifundefined
       {b@\@citeb}{{\bf ?}\@warning
       {Citation `\@citeb' on page \thepage \space undefined}}%
\hbox{\csname b@\@citeb\endcsname}}}{#1}}
\catcode`\@=12

\newcommand{\lsim}{\raisebox{-0.13cm}{~\shortstack{$<$ \\[-0.07cm] $\sim$}}~}

\newcommand{\beq}{\begin{eqnarray}}
\newcommand{\eeq}{\end{eqnarray}}

\title{SUSY-QCD corrections to MSSM Higgs boson production via gluon fusion}

\ShortTitle{SUSY-QCD corrections to MSSM Higgs boson production via gluon fusion}

\author{\speaker{Margarete M\"uhlleitner} \\
        Institute for Theoretical Physics, ITP (KIT)\\
        E-mail: \email{margarete.muehlleitner@particle.uni-karlsruhe.de}}

\author{Heidi Rzehak \\
        Institute for Theoretical Physics, ITP (KIT)\\
        E-mail: \email{hr@particle.uni-karlsruhe.de}}

\author{Michael Spira\\
        Paul Scherrer Institute, PSI (Villigen PSI)\\
        E-mail: \email{michael.spira@psi.ch}}

\abstract{In the MSSM scalar $h,H$ production is mediated by heavy
  quark and squark loops. The higher order QCD corrections have been 
 obtained some time ago and turned out to be large. The full SUSY QCD
 corrections have been obained recently including the full mass
 dependence of the loop particles. We describe our calculation and 
 present first numerical results. We also address the question of the
 proper treatment of the large gluino mass limit, {\it i.e.} the
 consistent decoupling of heavy gluino effects, and present the
 effective Lagrangian for decoupled gluinos.}

\FullConference{RADCOR 2009 - 9th International Symposium on Radiative Corrections (Applications of Quantum Field Theory to Phenomenology) ,\\
		 October 25 - 30 2009\\
		 Ascona, Switzerland}

\begin{document}

\section{Introduction}
One of the major goals at the LHC is the production of Higgs
boson(s) \cite{higgs}. In the Minimal Supersymmetric Extension of the
Standard Model (MSSM) two complex Higgs doublets are introduced to 
give masses to up- and down-type fermions
\cite{mssmhiggs}. After electroweak symmetry breaking there are five
physcial Higgs states, two CP-even neutral Higgs bosons $h,H$, one
neutral CP-odd Higgs state $A$ and two charged Higgs bosons
$H^\pm$. At tree level, the Higgs sector can be parameterized by two
independent parameters, the pseudoscalar Higgs boson mass $M_A$
and the ratio of the two vacuum expectation values (VEV) of the two
complex Higgs doublets, $\tan\beta=v_2/v_1$.  The Higgs couplings to
quarks and gauge bosons are modified with $\sin$ and $\cos$ of the
mixing angles $\alpha$ and $\beta$ with respect to the Standard Model (SM)
couplings, where $\alpha$ denotes the $h,H$ mixing angle. 
The bottom (top) Yukawa couplings are enhanced (suppressed) 
for large values of $\tan\beta$, so that top Yukawa
couplings play a dominant role at small and moderate values of
$\tan\beta$. 

At the LHC and Tevatron neutral Higgs bosons are copiously produced
via gluon fusion $gg \to h,H,A$, which is mediated in the case of
$h,H$ by (s)top and (s)bottom loops \cite{loproc}. The pure
QCD corrections to the (s)quark loops have been obtained including the
full Higgs and (s)quark mass dependences and increase the cross
sections by $\sim 100$\% \cite{nlocor}. This result can be
approximated by very heavy top (s)quarks with $\sim 20-30$\% accuracy
for $\tan\beta \lsim 5$ \cite{approx}. In this limit the
next-to-leading order (NLO) QCD \cite{nloqcd} and later the
next-to-next-to-leading order (NNLO) QCD corrections \cite{nnlo} have been
obtained, the latter leading to a moderate increase of
20-30\%. Finite top mass effects  at NNLO have been discussed in
\cite{finitetop}.
Finally, the estimate of the next-to-next-to-next-to-leading
order effects \cite{estimate} indicates improved perturbative convergence. The
full supersymmetric (SUSY) QCD corrections have been obtained in the limit of heavy
SUSY particle masses \cite{fullsusylim}
%\citer{fullsusylim1,fullsusylim4} 
and more
recently including the full mass dependence \cite{fullsusymass}. 
The electroweak loop effects have been calculated in \cite{ewcorr}.
%In Ref.\cite{fullsusylim1} the scalar Higgs coupling to gluons is given
%in the limit of large gluino masses for degenerate squarks, {\it i.e.}
%without mixing. The result develops a logarithmic singularity for
%large gluino masses thus apparently contradicting the
%Appelquist-Carazzone theorem \cite{appelquist}. This problem does not
%appear in the pseudoscalar Higgs case \cite{fullsusylim3}.
In this article we will describe in Section \ref{sec:hocalc} 
the calculation of the full SUSY-QCD
corrections in gluon fusion to $h,H$, and we will
present for the first time numerical
results for the total cross section. 
In Section \ref{sec:gludecoup} we will discuss the consistent derivation of the
effective Lagrangian for the scalar Higgs couplings to gluons after the
gluino decoupling.

\section{Gluon Fusion \label{sec:hocalc}}
\begin{figure}[t]
\begin{center}
\epsfig{figure=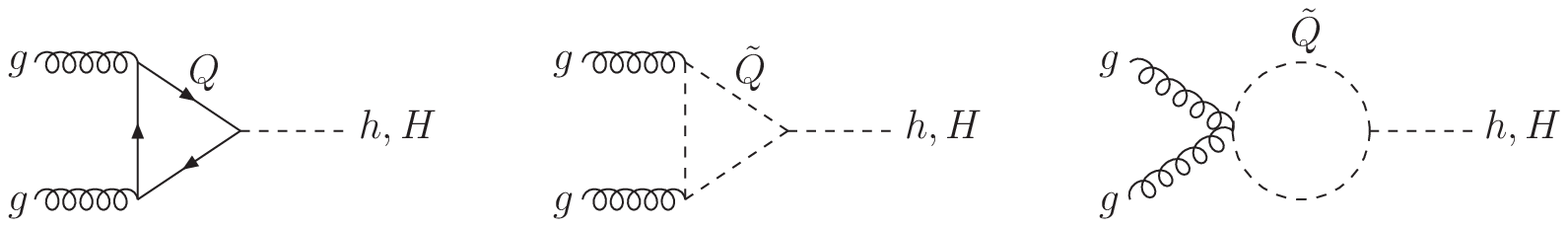,bbllx=100,bblly=605,bburx=600,bbury=690,width=0.8\textwidth,clip=}
\caption{Diagrams contributing to $gg\to h,H$ at leading order.}
\label{fig:loproc}
\end{center}
\end{figure}

\begin{figure}[hbtp]
\begin{center}
\epsfig{figure=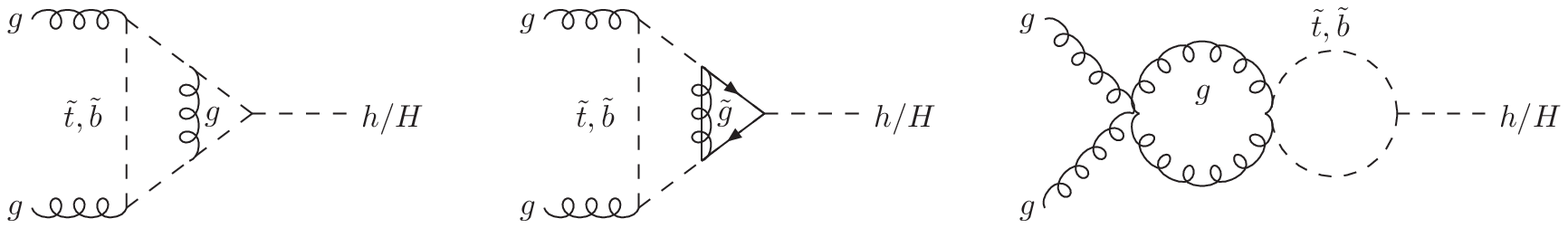,bbllx=0,bblly=584,bburx=600,bbury=720,width=0.9\textwidth,clip=}
\\[-1cm]
\caption{Some generic diagrams for the virtual NLO SUSY-QCD
  corrections to the squark contributions to the gluonic Higgs couplings.}
\label{fig:virtsusyqcd}
\end{center}
\end{figure}

\begin{figure}[hbtp]
\begin{center}
\epsfig{figure=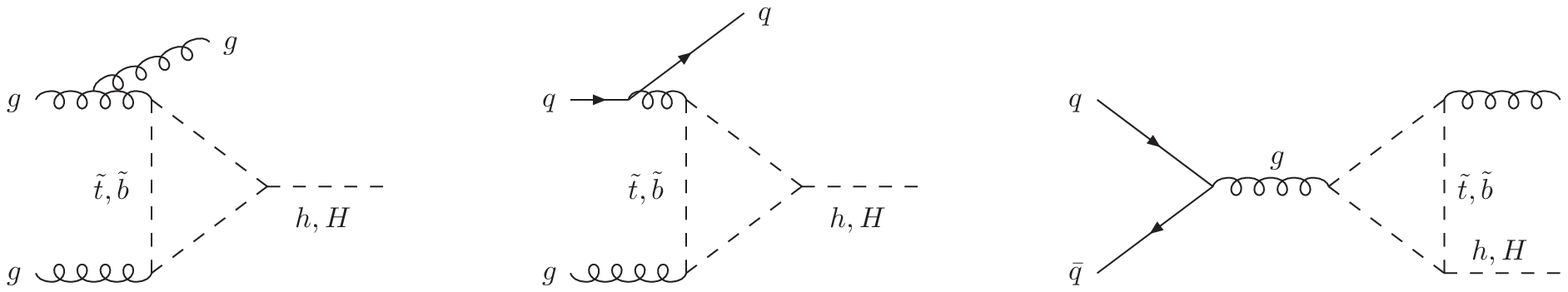,bbllx=0,bblly=570,bburx=600,bbury=720,width=0.8\textwidth,clip=}
\\[-1cm]
\caption{Typical diagrams for the real NLO QCD corrections to the squark contributions to the gluon fusion processes.}
\label{fig:realcor}
\end{center}
\end{figure}

\begin{figure}[ht]
\begin{center}
\begin{picture}(150,160)(0,0)
\put(-40,-90.0){\includegraphics{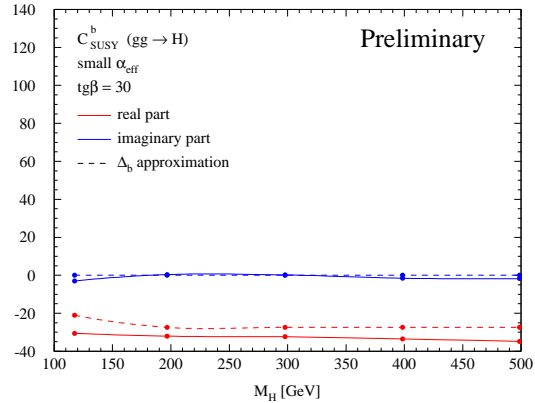}}
\put(110.0,130.0){\small Preliminary}
%\put(65.0,23.0){\large $\overline{\rm MS}$ $A_b(M_{\tildeg})$}
\end{picture}
%%\begin{picture}(40,40)(0,0)
%\epsfig{figure=cbsusy_ms.ps,bbllx=0,bblly=220,bburx=600,bbury=620,width=0.5\textwidth}
%%\special{psfile=cbsusy_on.ps hscale=30 vscale=30
%%                 angle=0 hoffset=0 voffset=0}
%%\end{picture}
\caption{The genuine SUSY QCD corrections 
normalized to the  LO bottom quark form factor. Real corrections: red
(light gray), virtual corrections: blue (dark gray), compared to the
$\Delta_b$ approximation (dashed lines). $A_b$ has been renormalized
in the $\overline{\mbox{MS}}$ scheme.}
\label{fig:totalnlo}
\end{center}
\end{figure}

At leading order (LO) the gluon fusion processes $gg\to h/H$ are mediated
by heavy quark and squark triangle loops, {\it cf.}
Fig.\ref{fig:loproc}, the latter contributing significantly for 
squark masses $\lsim 400$~GeV. The LO cross section in the
narrow-width approximation can be obtained from the $h/H$ gluonic decay
widths, \cite{loproc,georgi} 
\begin{eqnarray}
\sigma_{LO}(pp\to h/H) & = & \sigma^{h/H}_0 \tau_{h/H}\frac{d{\cal
L}^{gg}}{d\tau_{h/H}} \\
\sigma^{h/H}_0 & = & \frac{\pi^2}{8M_{h/H}^3}\Gamma_{LO}(h/H\to gg)
\nonumber \\
\sigma^{h/H}_0 & = & \frac{G_{F}\alpha_{s}^{2}(\mu_R)}{288 \sqrt{2}\pi} \
\left| \sum_{Q} g_Q^{h/H} A_Q^{h/H} (\tau_{Q})
+ \sum_{\widetilde{Q}} g_{\widetilde{Q}}^{h/H} A_{\widetilde{Q}}^{h/H}
(\tau_{\widetilde{Q}}) \right|^{2} \, , 
\end{eqnarray}
where $\tau_{h/H} = M_{h/H}^2/s$ with $s$ being the squared hadronic
c.m.\ energy and $\tau_{Q/\tilde{Q}}=4 m_{Q/\tilde{Q}}^2/M_{h/H}^2$. 
The LO form factors are given by
\beq
A_Q^{h/H}(\tau) & = & \frac{3}{2} \tau [1+(1-\tau)f(\tau)] \nonumber \\
A_{\tilde Q}^{h/H} (\tau) & = & -\frac{3}{4} \tau[1-\tau
f(\tau)] \\
f(\tau) & = & \left\{ \begin{array}{ll}
\displaystyle \arcsin^2 \frac{1}{\sqrt{\tau}} & \tau \ge 1 \\
\displaystyle - \frac{1}{4} \left[ \log \frac{1+\sqrt{1-\tau}}
{1-\sqrt{1-\tau}} - i\pi \right]^2 & \tau < 1
\end{array} \right. \nonumber\;.
\eeq
And the gluon luminosity at the factorization scale $\mu_F$ is defined as
\begin{displaymath}
\frac{d{\cal L}^{gg}}{d\tau} = \int_\tau^1 \frac{dx}{x}~g(x,\mu_F^2)
g(\tau /x,\mu_F^2) \, ,
\end{displaymath}
where $g(x,\mu_F^2)$ denotes the gluon parton density of the proton.
The NLO SUSY-QCD corrections consist of the virtual two-loop
corrections, {\it cf.} Fig.\ref{fig:virtsusyqcd}, and the real
corrections due to the radiation processes $gg\to gh/H, gq\to qh/H$
and $q\bar{q}\to gh/H$, {\it cf.} Fig.\ref{fig:realcor}.
The final result for the total hadronic cross sections can be split
accordingly into five parts,
\beq
\sigma(pp \rightarrow h/H+X) = \sigma^{h/H}_{0} \left[ 1+ C^{h/H}
\frac{\alpha_{s}}{\pi} \right] \tau_{h/H} \frac{d{\cal
L}^{gg}}{d\tau_{h/H}} + \Delta \sigma^{h/H}_{gg} + \Delta
\sigma^{h/H}_{gq} + \Delta \sigma^{h/H}_{q\bar{q}} \, .
\label{eq:glufuscxn}
\eeq
The strong coupling constant is renormalized in the $\overline{\rm MS}$
scheme, with the top quark and squark contributions decoupled from the
scale dependence. The quark and squark masses are renormalized
on-shell. The parton densities are defined in the $\overline{\rm MS}$
scheme with five active flavors, i.e. the top quark and the squarks are
not included in the factorization scale dependence. After
renormalization we are left with collinear divergences in the sum of
the virtual and real corrections which are absorbed in the
renormalization of the parton density functions, so that the result
Eq.~(\ref{eq:glufuscxn}) is finite and depends on the
renormalization and factorization scales $\mu_R$ and $\mu_F$,
respectively. The  natural scale choices turn out to be
$\mu_R=\mu_F\sim M_{h/H}$. 
The numerical results are presented for the modified small
$\alpha_{eff}$ scenario \cite{alphaeff}, defined by the
following choices of MSSM parameters [$m_t=172.6$~GeV],
\beq
\begin{array}{llllll}
M_{\tilde{Q}} &=& 800\;\mbox{GeV} & \qquad \tan\beta &=& 30 \\
M_{\tilde{g}} &=& 1000\;\mbox{GeV} & \qquad \mu &=& 2\;\mbox{TeV} \\
M_2 &=& 500\;\mbox{GeV} & \qquad A_b = A_t &=& -1.133\;\mbox{TeV} \;.
\end{array}
\eeq
In this scenario the squark masses amount to
\beq
\begin{array}{llllll}
m_{\tilde{t}_1} &=& 679\;\mbox{GeV} & \qquad m_{\tilde{t}_2} &=&
935\;\mbox{GeV} \\
m_{\tilde{b}_1} &=& 601\;\mbox{GeV} & \qquad m_{\tilde{b}_2} &=&
961\;\mbox{GeV} \;.
\end{array}
\eeq
Fig. \ref{fig:totalnlo} displays the genuine SUSY QCD corrections 
normalized to the  LO bottom quark form factor, {\it i.e.}
$A_b^{h/H} (\tau_b) \to A_b^{h/H} (\tau_b) (1+C^b_{SUSY}
\frac{\alpha_s}{\pi})$. The corrections can be sizeable, but can be
described reasonably with the usual $\Delta_b$ approximation
\cite{guasch}, if $A_b$ is renormalized in the $\overline{\mbox{MS}}$ scheme.

\section{Decoupling of the Gluinos \label{sec:gludecoup}}
In this section we will address the limit of heavy quark, squark and
gluino masses, where in addition the gluinos are much heavier than the quarks and
squarks. For the derivation of the effective Lagrangian
for the scalar Higgs couplings to gluons we analyze the relation
between the quark Yukawa coupling $\lambda_Q$ and the Higgs coupling
to squarks $\lambda_{\tilde{Q}}$ in the limit of large gluino
masses. We define these couplings at leading order in the case of
vanishing mixing,
\beq
\lambda_Q = g_Q^{\cal H} \frac{m_Q}{v}\; , \qquad \lambda_{\tilde{Q}} = 2
g_Q^{\cal H} \frac{m_Q^2}{v} = \kappa \lambda_Q^2\; , \qquad 
\mbox{with }
\kappa = 2 \frac{v}{g_Q^{\cal H}} \;,
\label{eq:relation}
\eeq
where $g_Q^{\cal H}$ denotes the normalization factor of the MSSM
Higgs couplings to quark pairs with respect to the SM. In the
following we will sketch how the modified relation between these
couplings for scales {\it below} the gluino mass $M_{\tilde{g}}$ is
derived. For details, see Ref.~\cite{gldecoup}. We
start with the unbroken relation between the running
$\overline{\mbox{MS}}$ couplings of Eq.~(\ref{eq:relation}) and the
corresponding renormalization group equations (RGE) for scales
{\it above} $M_{\tilde{g}}$. If the scales decrease {\it below}
$M_{\tilde{g}}$ the gluino decouples from the RGEs leading to modified
RGEs which are different for the two couplings $\lambda_{\tilde{Q}}$
and $\kappa \lambda_Q^2$ so that the two couplings deviate for scales
below $M_{\tilde{g}}$. The proper matching at the gluino mass scale
yields a finite threshold contribution for the evolution from the
gluino mass scale to smaller scales, while the logarithmic structure of
the matching relation is given by the solution of the RGEs {\it below}
$M_{\tilde{g}}$. 
%The matching relations can be obtained from the
%gluino contributions for heavy gluino masses in the limit of vanishing
%external momentum transfers \cite{momsubs}. 
In order to decouple consistently the gluino from the RGE for gluino
mass scales large compared to the chosen renormalization scale, a
momentum substraction of the gluino contribution for vanishing
momentum transfer has to be performed \cite{momsubs}. We refer the reader
to \cite{gldecoup} for details and give here directly the result for
the modified relation between the quark Yukawa coupling and the
effective Higgs coupling to squarks taking into account the proper gluino
decoupling:
\beq
2 g_Q^{\cal H} \frac{m_Q^2}{v} = \bar \lambda_{\tilde Q,MO}(m_{\tilde Q}) \left\{ 1 +
C_F\frac{\alpha_s}{\pi} \left(\log\frac{M_{\tilde{g}}^2}{m_{\tilde Q}^2} +
\frac{3}{2}\log\frac{m_{\tilde Q}^2}{m_Q^2} + \frac{1}{2} \right)
\right\} \;,
\eeq
where $m_Q$ is the pole mass and $MO$ denotes the momentum substracted
coupling, which is taken at the squark mass scale, which is the
proper scale choice of 
the effective Higgs coupling to squarks and which is relevant for an
additional large gap between the quark and squark masses. 

Taking into account the radiative corrections to the relation between the effective
couplings after decoupling the gluinos leads to the following
effective Lagrangian in the limit of heavy squarks and quarks,
\begin{equation}
{\cal L}_{eff} = \frac{\alpha_s}{12\pi} G^{a\mu\nu} G^a_{\mu\nu}
\frac{\cal H}{v} \left\{\sum_Q g_Q^{\cal H}
\left[1+\frac{11}{4}\frac{\alpha_s}{\pi}\right] + \sum_{\tilde Q}
\frac{g_{\tilde Q}^{\cal H}}{4} \left[1 + C_{SQCD}
\frac{\alpha_s}{\pi}\right] + {\cal O}(\alpha_s^2) \right\} \; ,
\end{equation}
where $g_{\tilde Q}^{\cal H}=v\bar \lambda_{\tilde Q,MO}(m_{\tilde Q}) /
m_{\tilde Q}^2$. The cofficient $C_{SQCD}$ is given by
\beq
C_{SQCD} = \frac{37}{6} \;.
\eeq
It is well-defined in the limit of large gluino masses and thus
fulfills the constraint of the Appelquist--Carazzone decoupling theorem
\cite{appelquist}. 

\section{Conclusions}
We have presented first results for the NLO SUSY QCD corrections to
gluon fusion into CP-even MSSM Higgs bosons, including the full mass
dependence of the loop particles. The genuine SUSY-QCD corrections can
be sizeable. We furthermore demonstrated, that the gluino
contributions can be decoupled in the large $M_{\tilde{g}}$ limit in
accordance with the Appelquist-Carazzone theorem.


\begin{thebibliography}{99}
\bibitem{higgs} P.W.~Higgs, Phys. Lett. {\bf 12} (1964) 132;
  Phys. Rev. Lett. {\bf 13} (1964) 508 and Phys. Rev. {\bf 145} (1966)
  1156; F.~Englert and R.~Brout, Phys. Rev. Lett. {\bf 13} (1964) 321;
  G.S.~Guralnik, C.R.~Hagen and T.W.~Kibble, 
  Phys. Rev. Lett. {\bf 13} (1964) 585.

\bibitem{mssmhiggs} P.~Fayet, Nucl. Phys. {\bf B90} (1975) 104,
  Phys. Lett. {\bf B64} (1976) 159 and
  Phys. Lett. {\bf B69} (1977) 489; S.~Dimopoulos and H.~Georgi,
  Nucl. Phys. {\bf B193} (1981) 150; N.~Sakai, Z. Phys. {\bf C11}
  (1981) 153; K.~Inoue et al.,
%, A.~Kakuto, H.~Komatsu and S.~Takeshita,
  Prog. Theor. Phys. {\bf 67} (1982) 1889, Prog. Theor. Phys. {\bf 68}
  (1982) 927 [Erratum-ibid. {\bf 70} (1983) 330] and
  Prog. Theor. Phys. {\bf 71} (1984) 413. 

\bibitem{loproc} M.~Spira, Fortsch. Phys. {\bf 46} (1998) 203;
  A.~Djouadi, Phys. Rept. {\bf 457} (2008) 1 and {\bf 459} (2008) 1.

\bibitem{nlocor} D.~Graudenz, M.~Spira and P.M.~Zerwas,
  Phys. Rev. Lett. {\bf 70} (1993) 1372; M.~Spira et al.,
%, A.~Djouadi, D.~Graudenz  and P.M.~Zerwas, 
  Phys. Lett. {\bf B318} (1993) 347 and
  Nucl. Phys. {\bf B453} (1995) 17; C.~Anastasiou et al.,
%, S.~Beerli, A.~Daleo and Z.~Kunszt, 
  JHEP {\bf 0701} (2007) 082; U.~Aglietti et al., JHEP {\bf 0701}
  (2007) 021; R.~Bonciani et al.,
%, G.~Degrassi and A.~Vicini, 
JHEP {\bf 0711} (2007) 095; M.~M\"uhlleitner and
  M.~Spira, Nucl. Phys. {\bf B790} (2008) 1.

\bibitem{approx} M.~Kr\"amer, E.~Laenen and M.~Spira, Nucl. Phys. {\bf
  B511} (1998) 523; M.~Spira, arXiv:hep-ph/9703355.

\bibitem{nloqcd} A.~Djouadi, M.~Spira and P.M.~Zerwas,
  Phys. Lett. {\bf B264} (1991) 440; S.~Dawson, Nucl. Phys. {\bf B359}
  (1991) 283; R.P.~Kauffman and W.~Schaffer, Phys. Rev. {\bf D49}
  (1994) 551; S.~Dawson and R.~Kauffman, Phys. Rev. {\bf D49} (1994)
  2298; S.~Dawson, A.~Djouadi and M.~Spira, Phys. Rev. Lett. {\bf 77}
  (1996) 16.

\bibitem{nnlo} R.V.~Harlander and W.B.~Kilgore, Phys. Rev. Lett. {\bf
    88} (2002) 201801 and JHEP {\bf 0210} (2002) 017; C.~Anastasiou
  and K.~Melnikov, Nucl. Phys. {\bf B646} (2002) 220 and
  Phys. Rev. {\bf D67} (2003) 037501; V.~Ravindran, J.~Smith and
  W.L.~van Neerven, Nucl. Phys. {\bf B665} (2003) 325.

\bibitem{finitetop} R.V.~Harlander and K.J.~Ozeren, Phys. Lett. {\bf B679}
  (2009) 467 and JHEP {\bf 0911} (2009) 088; A.~Pak, M.~Rogal and
  M.~Steinhauser, Phys. Lett. {\bf B679} (2009) 473 and
  arXiv:0911.4662; R.V.~Harlander et al., arXiv:0912.2104 [hep-ph].

\bibitem{estimate} S.~Catani et al.,
%, D.~de Florian, M.~Grazzini and P.~Nason,
  JHEP {\bf 0307} (2003) 028; S.~Moch and A.~Vogt, Phys. Lett. {\bf
    B631} (2005) 48; V.~Ravindran, Nucl. Phys. {\bf B746} (2006) 58
  and Nucl. Phys. {\bf B752} (2006) 173.

\bibitem{fullsusylim} R.V.~Harlander and M.~Steinhauser, Phys.\ Lett.\
  {\bf B574} (2003) 258, Phys.\ Rev.\ {\bf D68} (2003) 111701 and JHEP
  {\bf 0409} (2004) 066; R.V.~Harlander and F.~Hofmann, JHEP {\bf
    0603} (2006) 050; G.~Degrassi and P.~Slavich, Nucl.\ Phys.\  {\bf
    B805} (2008) 267.

\bibitem{fullsusymass} 
 C.~Anastasiou, S.~Beerli and A.~Daleo, Phys.\ Rev.\ Lett.\  {\bf
   100} (2008) 241806.

\bibitem{ewcorr}  G.~Degrassi and F.~Maltoni, Phys. Lett. {\bf
    B600} (2004) 255; U.~Aglietti et al., hep-ph/0610033; S.~Actis et al.,
%, R.~Bonciani, G.~Degrassi and A. Vicini, 
%, G.Passarino, C.~Sturm, and S.Uccirati, 
  Phys. Lett. {\bf B670} (2008) 12; C.~Anastasiou et al.,
%, R.Boughezal and F.Petriello, 
JHEP {\bf 0904} (2009) 003. 

%\bibitem{fullsusylim1} R.V.~Harlander and M.~Steinhauser,
% Phys.\ Lett.\ {\bf B574} (2003) 258.
%
%\bibitem{fullsusylim2} R.~Harlander and M.~Steinhauser,
% Phys.\ Rev.\ {\bf D68} (2003) 111701 and JHEP {\bf 0409} (2004) 066.
%
%\bibitem{fullsusylim3} R.V.~Harlander and F.~Hofmann,
% JHEP {\bf 0603} (2006) 050.
%
%\bibitem{fullsusylim4}
%  G.~Degrassi and P.~Slavich, Nucl.\ Phys.\  {\bf B805} (2008) 267.

\bibitem{georgi} H.M.~Georgi, S.L.~Glashow, M.E.~Machacek and
  D.V.~Nanopoulos, Phys. Rev. Lett. {\bf 40} (1978) 692.

\bibitem{alphaeff} M~Carena et al.,
%, S.~Heinemeyer, C.E.M.~Wagner and G.~Weiglein, 
Eur. Phys. J. {\bf C26} (2003) 601.

\bibitem{guasch} M.~Carena et al.,
%, D.~Garcia, U.~Nierste and C.E.~Wagner,
  Nucl. Phys. {\bf B577} (2000) 88; J.~Guasch et al.,
%, P.~H\"afliger and M.~Spira,
  Phys. Rev. {\bf D68} (2003) 11501.

\bibitem{gldecoup} M.~M\"uhlleitner, H.Rzehak and M.Spira, JHEP {\bf
    0904} (2009) 023.

\bibitem{momsubs} J. C. Collins, F. Wilczek and A. Zee,
  Phys. Rev. {\bf D18} (1978) 242.

\bibitem{appelquist}   T.~Appelquist and J.~Carazzone,
 Phys.\ Rev.\  {\bf D11} (1975) 2856.

\end{thebibliography}
\end{document}